\documentclass[12pt]{article}
\usepackage{epsfig}
\usepackage{psfrag,pstricks,pst-node,pst-text,pst-3d}
\usepackage[latin1]{inputenc}
\usepackage{rotating}
\usepackage{graphicx}
\usepackage{amsbsy,amsmath,amsfonts,amssymb}
\usepackage{axodraw}

\def\beq{\begin{equation}}
\def\eeq#1{\label{#1}\end{equation}}
\def\eeqn{\end{equation}}

\def\beqa{\begin{eqnarray}}
\def\eeqa#1{\label{#1}\end{eqnarray}}
\def\eeqan{\end{eqnarray}}

\let\bar=\overbar

\def\Dslash{\not{\hbox{\kern-4pt $D$}}}
\def\dslash{\not{\hbox{\kern-2pt $\del$}}}

\def\msb{{\bar{\ssstyle M \kern -1pt S}}}

\newcommand{\xilampi}{\Xi^0 \to \Lambda \pi^0}

\newcommand{\xisigenu}{\Xi^0 \to \Sigma^+ e^- \bar{\nu}}
\newcommand{\axisigenu}{\overline{\Xi^0} \to \overline{\Sigma^+} e^+ \nu}
\newcommand{\xisigmunu}{\Xi^0 \to \Sigma^+ \mu^- \bar{\nu}}
\newcommand{\xilamgam}{\Xi^0 \to \Lambda \gamma}
\newcommand{\xisiggam}{\Xi^0 \to \Sigma^0 \gamma}
\newcommand{\Br}{{\text{Br}}}
\def\Journal#1#2#3#4{{#1} {\bf #2} (#4) #3}

\def\Title#1{\begin{center} {\Large {\bf #1} } \end{center}}

\begin{document}

\Title{New Results on $\Xi^0$ Hyperon Decays}

\begin{center}{\large \bf Contribution to the proceedings of HQL06,\\
Munich, October 16th-20th 2006}\end{center}

\bigskip\bigskip


\begin{raggedright}  

{\it Rainer Wanke\index{Wanke, R.}\\
Institut f\"ur Physik,\\
Universit\"at Mainz\\
D-55099 Mainz, GERMANY}
\bigskip\bigskip
\end{raggedright}

\section{Introduction}

Hyperons have been among the first non-stable elementary particles discovered
and are known for more than 50 years now.
In spite of this, the physics interest in hyperons is by far not yet exhausted,
several important aspects are still under intense investigation.
At first, as hyperons only differ by one or two strange-quarks from 
proton and neutron,
they are ideal playgrounds to study $SU(3)_f$ symmetry breaking.
Secondly, studying their decays offers unique opportunities 
to understand baryon structure and decay mechanisms.
Last, but not least, by comparing hyperon decays to neutron decay, it is possible to
measure the CKM matrix parameter $|V_{us}|$ in a complementary way
with respect to kaon decays.

However, despite of their interest, most previous measurements date back to the 1960's
and 70's, in particular those on the neutral $\Xi^0$ hyperon.
Only recently new measurements have been performed, with much increased statistics and leading to
a series of new results.

The two main new experiments 
on $\Xi^0$ decays are the KTeV experiment at Fermilab and NA48/1 at the CERN SPS.
Both experiments were designed to measure neutral kaon decays
and profit from the fact, that the $\Xi^0$ lifetime and decay length are of the same order
as those of the $K^0_S$ meson.
Most of the new results are coming from NA48/1. This experiment was performed in the year
2002 to measure specifically rare $K^0_S$ and neutral hyperon decays.
During the data taking period, in total more than 2 billion of $\Xi^0$ decays took place 
in the fiducial detector volume, providing enough statistics to precisely measure also
very rare $\Xi^0$ decays.

\section{Measurement of the $\Xi^0$ Lifetime}

The $\Xi^0$ lifetime, i.e.\ its total decay rate, is a key parameter for
interpreting other measurements on $\Xi^0$ decays. In particular for the determination
of the CKM matrix element $|V_{us}|$, as reported in the following section,
the $\Xi^0$ lifetime is a direct input parameter.
However, the precision on the $\Xi^0$ lifetime is very poor. 
The current world average $\tau_{\Xi^0} = (2.90 \pm 0.03) \times 10^{-10}$~s~\cite{bib:pdg06}
has less than $3\%$ accuracy, and the last measurement dates back to 1977. 

\begin{figure}[thb]
  \begin{center}
    \epsfig{file=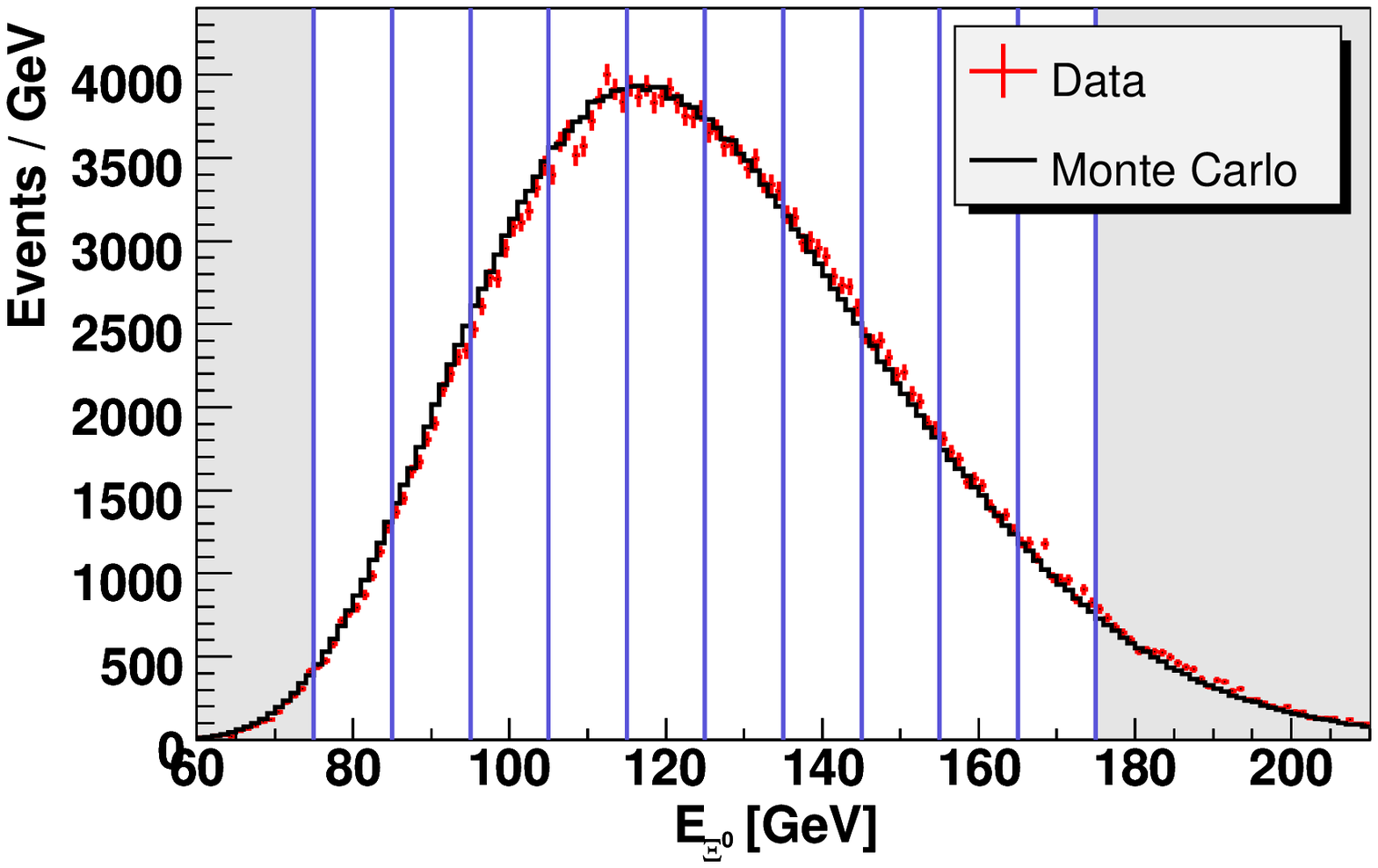,width=0.53\linewidth} 
    \epsfig{file=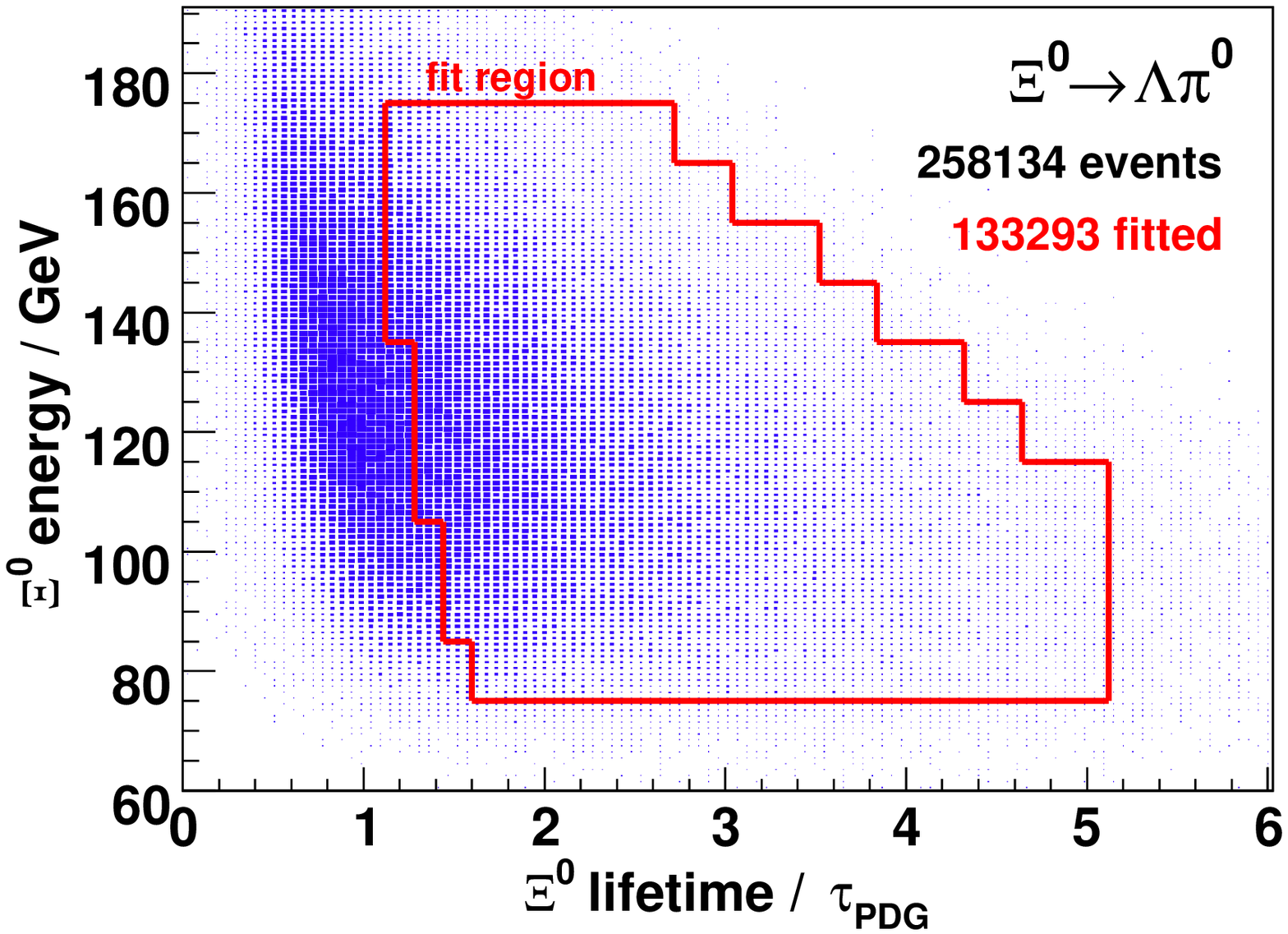,width=0.46\textwidth}
    \caption{Left: Energy spectrum of $\xilampi$ data and Monte Carlo events used in the lifetime fit.
             Right: Energy versus proper lifetime of the selected events. Indicated
             is the region used in the fit.}             
    \label{fig:xi0energy}
  \end{center}
\end{figure}

For a new, much more precise measurement the NA48/1 Collaboration has used 
$\xilampi$ decays taken with a minimum bias trigger. In total about 260~000 decays
have been selected with completely negligible background.
The energy spectrum of the selected events is shown in Fig.~\ref{fig:xi0energy} (left).
To be insensitive to the small residual differences between data and simulation
in the spectrum, the lifetime distributions are split Ted into 10 separate bins of energy, indicated by vertical lines in
Fig.~\ref{fig:xi0energy}.
The fit region is shown in Fig.~\ref{fig:xi0energy} (right). To avoid effects from
the vertex resolution differences between data and simulation at the beginning of the 
decay region, the fit region is well separated from the final collimator.
This requirement rejects about half of the statistics, leaving 133~293 events to enter the fit.
The fit to the lifetime is performed using the least-squares method, with
the normalisations in each energy bin left as free parameters.

\begin{figure}[thb]
  \begin{center}
    \epsfig{file=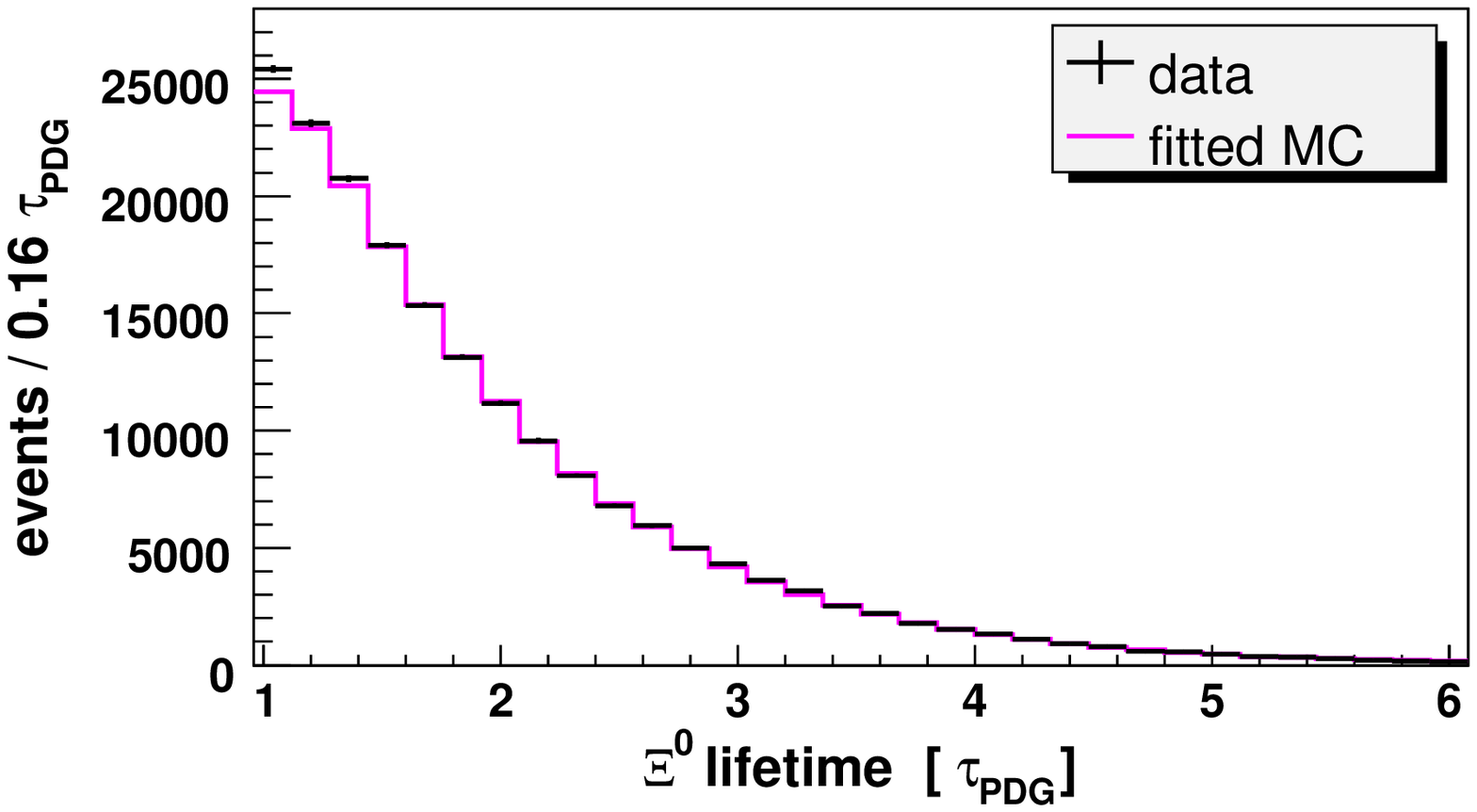,width=0.49\linewidth}
    \hspace*{\fill}
    \epsfig{file=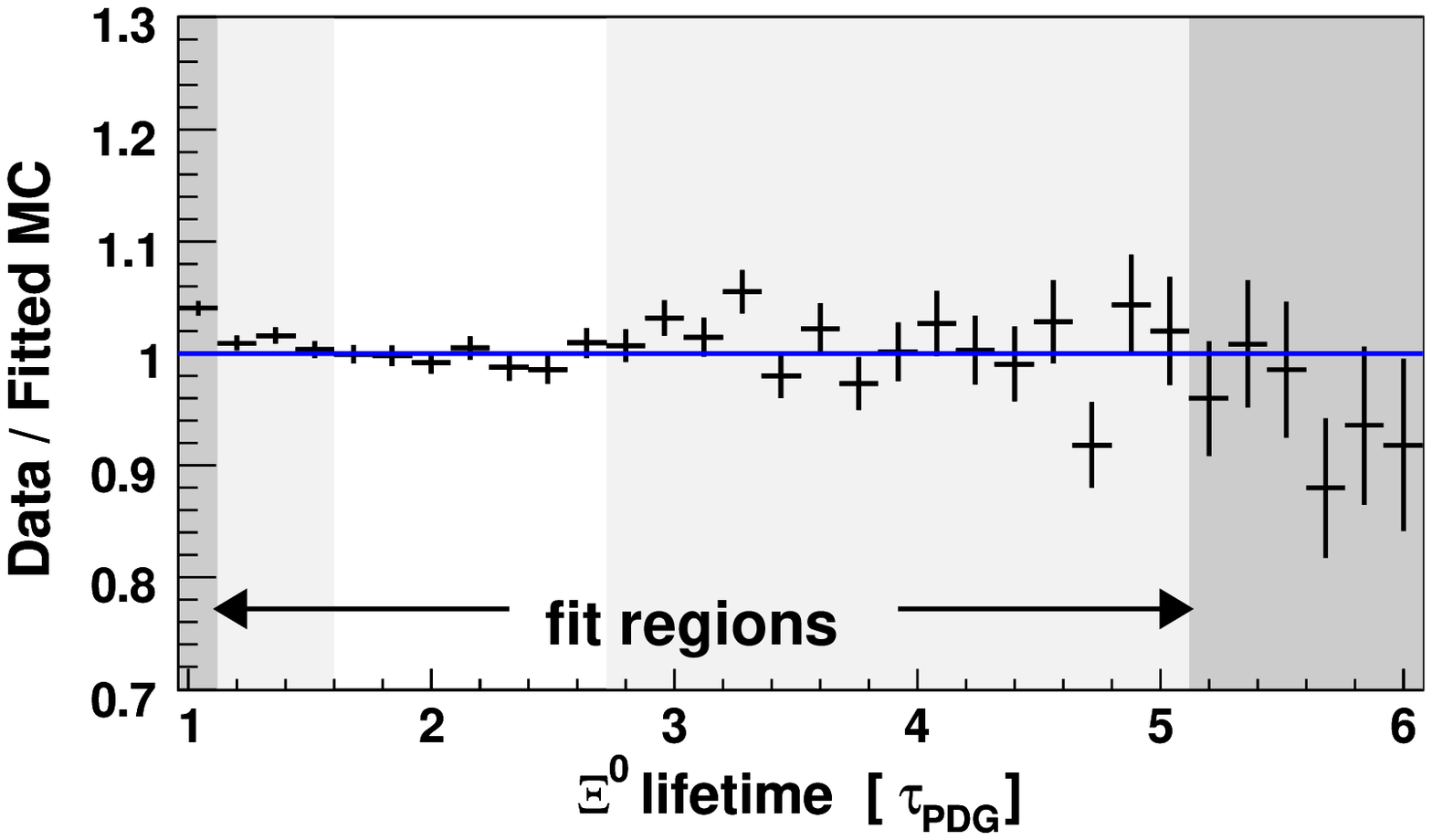,width=0.49\linewidth}
    \caption{Fit of the $\Xi^0$ lifetime. Left: Comparison of data and fitted Monte Carlo simulation
            as function of the current PDG lifetime $\tau_\text{PDG} = 2.90 \times 10^{-10}$~s~\cite{bib:pdg06}.
            Right: Fit residuals. The white (light grey) regions indicate the regions where the fit has been performed
            for all (some) energy bins.}
    \label{fig:lifetimefit}
  \end{center}
\end{figure}

The fit result, integrated over the separate energy bins, is shown in Fig~\ref{fig:lifetimefit}.
The NA48/1 collaboration obtains as preliminary result
\begin{equation}
\tau_{\Xi^0} = (3.082 \pm 0.013_\text{stat} \pm 0.012_\text{syst}) \cdot 10^{-10} \: \text{s},
\end{equation}
with the systematics being dominated by uncertainties of the detector acceptance,
the nominal $\Xi^0$ mass, and the $\Xi^0$ polarisation.
The result is about 2 standard deviations above the current 
world average and five times more precise.

\section{Measurement of the $\Xi^0$ Beta Decay \\ and Determination of $|V_{us}|$}

The $\Xi^0$ beta decay $\xisigenu$ is similar 
to the neutron beta decay with the down-quarks exchanged by strange-quarks.
The decay therefore is well suited for a measurement of the CKM parameter
$|V_{us}|$ complementary to the usual determination from kaon semileptonic decays.
A previous measurement has been published by the KTeV Collaboration, based on 
176~events~\cite{bib:Xi0Sigenu_KTeV_pub}.
An additional preliminary KTeV result, based on 626 events, has been presented
on conferences~\cite{bib:Xi0Sigenu_KTeV_prel}.

\begin{figure}[htb]
  \begin{center}
    \epsfig{file=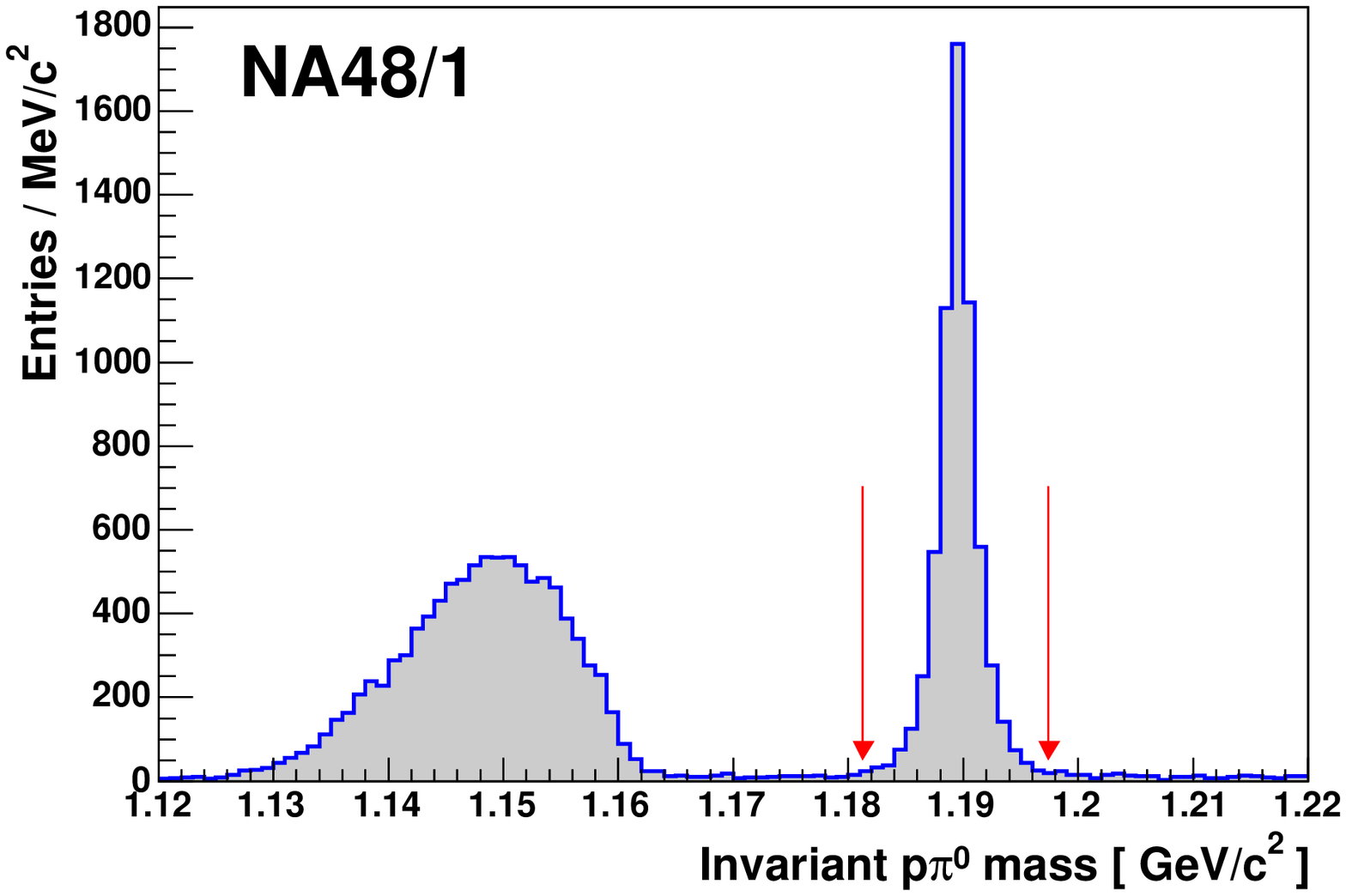,width=0.6\linewidth}
    \caption{Invariant $p\pi^0$ mass distribution from NA48/1
             $\xisigenu$ decays.}
    \label{fig:xisigenu}
  \end{center}
\end{figure}

With much larger statistics, NA48/1 has now performed a precise measurement of the
$\Xi^0$ beta decay. 
Since $\Xi^0$ semileptonic decays are the only source of $\Sigma^+$ in the neutral beam,
it is sufficient to select $\Sigma^+ \to p \pi^0$ events and require an additional
electron. The invariant $p \pi^0$ mass distribution of the accepted events is shown
in Fig.~\ref{fig:xisigenu}, yielding 6316 $\xisigenu$ candidates 
with an estimated background contamination of about $3\%$ mainly from in-time
accidental overlaps.

Normalising to the abundant decay $\xilampi$, NA48/1 obtains~\cite{bib:Xi0Sigenu_NA48}
\begin{equation}
\Br(\xisigenu) = ( 2.51 \pm 0.03_\text{stat} \pm 0.09_\text{syst}) \times 10^{-4}
\end{equation}
The systematic uncertainty is dominated by the statistical uncertainty of the
determination of the trigger efficiency ($\pm 2.2\%$). Further contributions
are the limited knowledge of the decay form factors $g_1$ and $f_2$ ($\pm 1.6\%$),
the detector acceptance, the $\Xi^0$ polarisation, and the normalisation
($\pm 1.0\%$ each).

In addition to the $\Xi^0$ decay, 555 $\axisigenu$ candidates were selected with a background of
about 136 events, yielding a branching fraction of
$\Br(\axisigenu) = ( 2.55 \pm 0.14_\text{stat} \pm 0.10_\text{syst}) \times 10^{-4}$
in perfect agreement with the one obtained from the $\Xi^0$ decay.

Using the combined result $\Br(\xisigenu) = (2.51 \pm 0.09) \times 10^{-4}$ and the
new preliminary value for the $\Xi^0$ lifetime, the partial decay width
is $\Gamma(\xisigenu) = (8.14 \pm 0.29) \times 10^5$~s$^{-1}$, from which
the CKM matrix element $|V_{us}|$ can be determined.
The semileptonic decay rate is given by~\cite{bib:garcia}
\begin{eqnarray*}
\Gamma & = & G_F^2 \, |V_{us}|^2 \, \frac{\Delta m^5}{60 \pi^3} \, (1+\delta_\text{rad}) \\*[1.5mm]
       &  &  \!\!\!\!\!\!  \times \left[ \left( 1 - \frac{3}{2} \, \beta \right) \left( |f_1|^2 + |g_1|^2 \right)
                           + \frac{6}{7} \, \beta^2 \left( |f_1|^2 + 2 |g_1|^2 +\text{Re}(f_1 f_2^\star) 
                                                                                  + \frac{2}{3} \, |f_2^2| \right)
                           + \delta_{q^2} \right],
\end{eqnarray*}
with $\Delta m = m_{\Xi^0} - m_{\Sigma^+}$~\cite{bib:pdg06}, $\beta = \Delta m / m_{\Xi^0}$,
the radiative corrections $\delta_\text{rad}$,
and $\delta_{q^2}(f_1,g_1)$ taking into transfer momentum dependence of $f_1$ and $g_1$~\cite{bib:garcia}. 
The form factor ratios $g_1/f_1$ and $f_2/f_1$
have been measured by the KTeV Collaboration~\cite{bib:Xi0FF_KTeV}.
Neglecting $SU(3)$ breaking corrections for $f_1$, a value of 
\begin{equation}
|V_{us}| = 0.203 \pm 0.004_\text{exp}  \begin{array}{l} +0.022 \\ -0.027 \end{array}_\text{form factors} 
\end{equation}
is found, in good agreement with the value of $0.226 \pm 0.002$ obtained from kaon decays~\cite{bib:pdg06},
but still large uncertainties from the form factor measurement.

\section{First Measurements of the Decay $\Xi^0 \to \Sigma^+ \mu \nu_\mu$}

\begin{figure}[thb]
  \begin{center}
    \epsfig{file=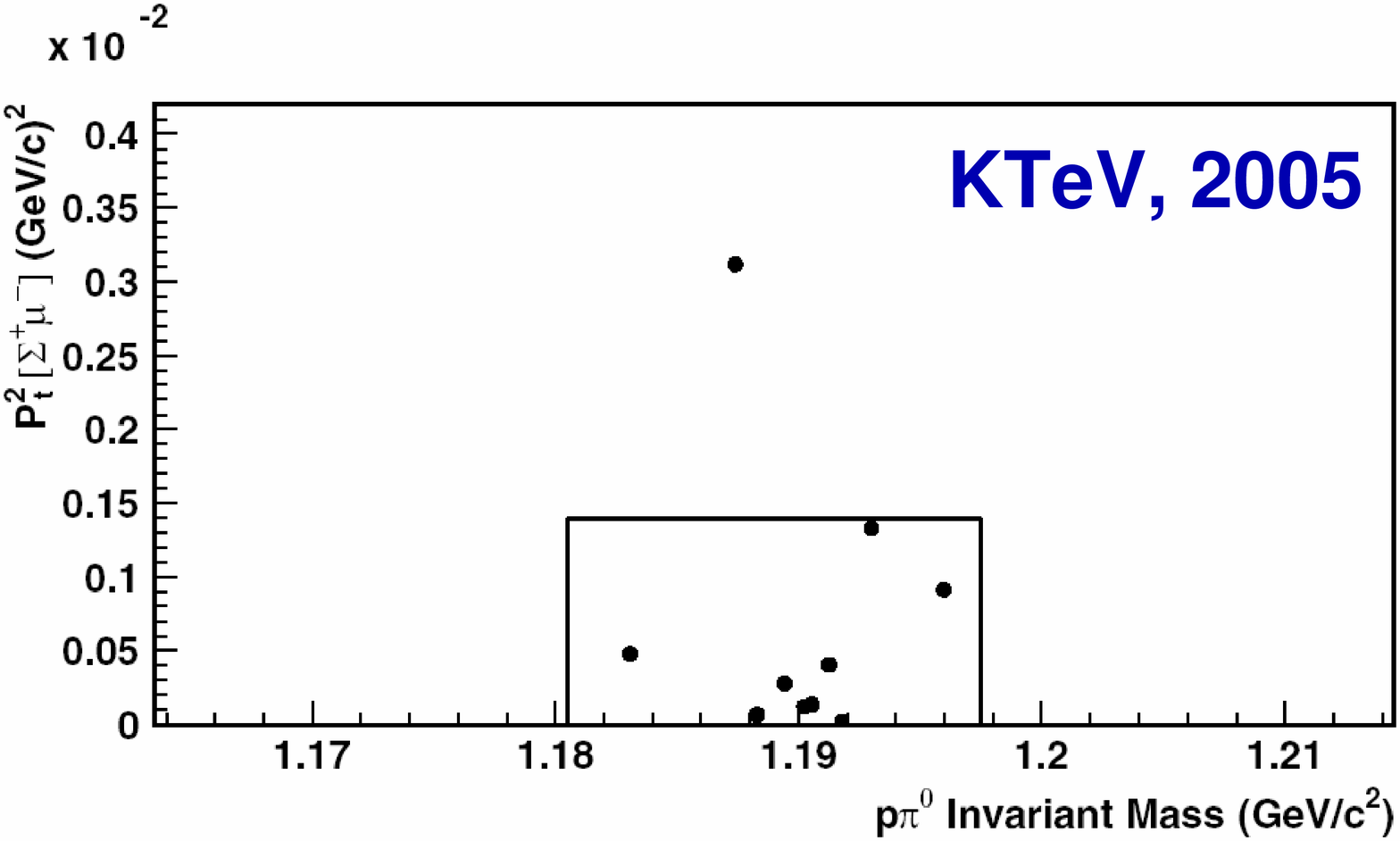,width=0.52\linewidth}
    \hspace*{\fill}
    \epsfig{file=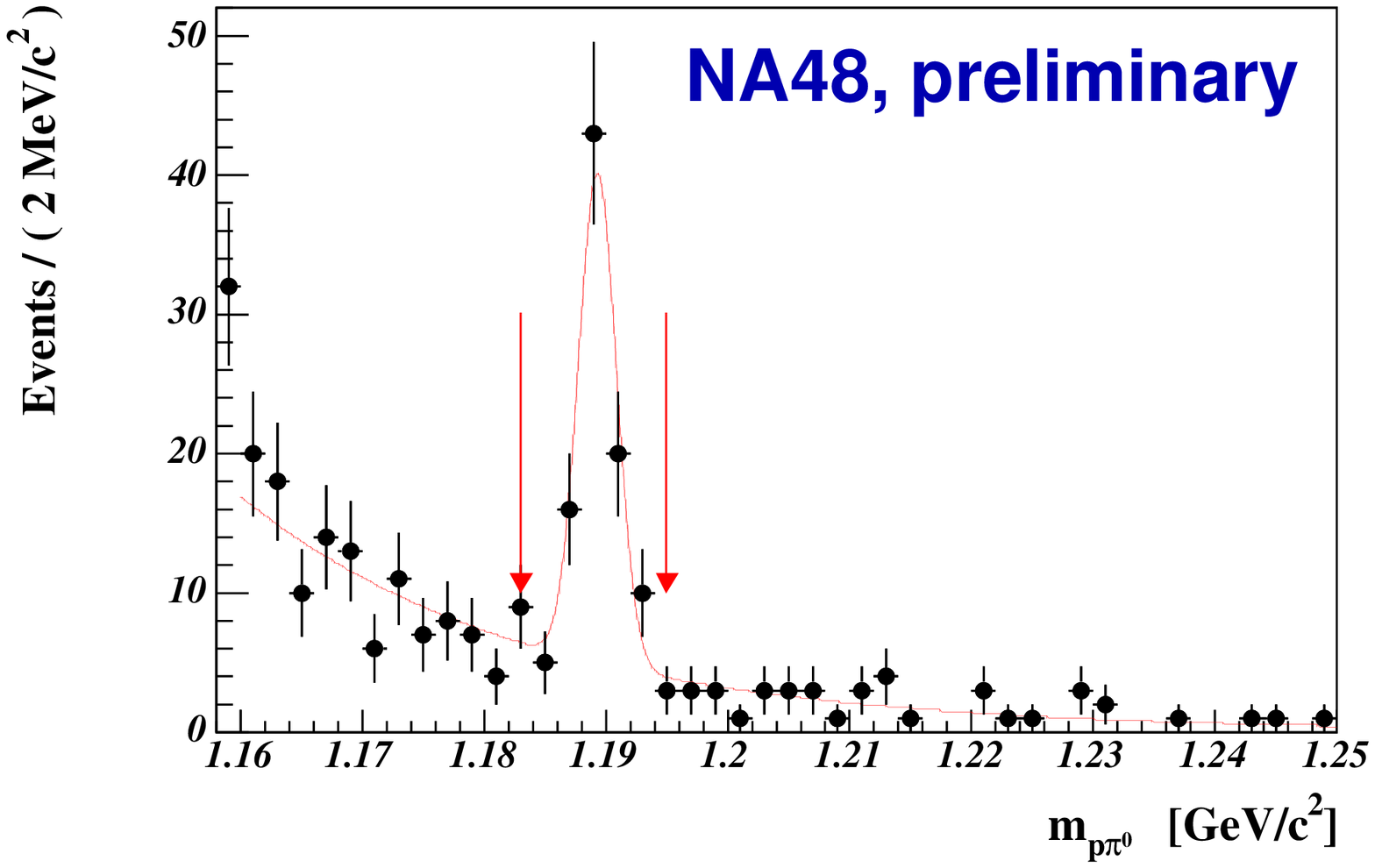,width=0.46\linewidth}
    \caption{Invariant $p \pi^0$ mass distributions of $\xisigmunu$ decays measured 
             by KTeV (left) and NA48/1 (right).}
    \label{fig:xisigmunu}
  \end{center}
\end{figure}

Because of the small available phase space, the semimuonic $\Xi^0$ decay is
highly suppressed with respect to its semielectronic counterpart.
The first observation of the decay was done by the KTeV Collaboration
in 2005~\cite{bib:Xi0Sigmunu_KTeV}. 
They observe 8 signal candidates over negligible background (Fig.~\ref{fig:xisigmunu} (left))
from which they obtain a branching fraction of
$\Br(\xisigmunu) = (4.7^{\,+2.0}_{\,-1.4} \pm 0.8 ) \times 10^{-6}$.

More recently, the NA48/1 Collaboration reported a preliminary measurement, based on 
99~signal candidates including about 30 background events (Fig.~\ref{fig:xisigmunu} (right)),
leading to a value of $\Br(\xisigmunu) = (2.2 \pm 0.3 \pm 0.2 ) \times 10^{-6}$.

\section{Weak Radiative $\Xi^0$ Decays}

\begin{figure}[htb]
  \begin{center}
    \epsfig{file=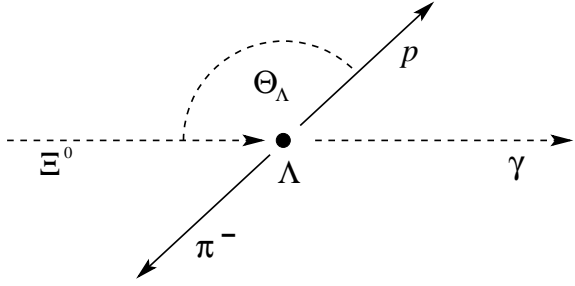,width=0.5\linewidth}
    \caption{Definition of the angle $\Theta$ between the proton and the incoming $\Xi^0$ in the Lambda rest frame.}
    \label{fig:Xi0Lamgamsketch}
  \end{center}
\end{figure}

Up to this day, weak radiative hyperon decays as $\xilamgam$ and $\xisiggam$ are still barely understood. 
Several competing theoretical models exist, which give very different predictions.
An excellent experimental parameter to distinguish between models is the decay asymmetry $\alpha$
of these decays. It is defined as 
\begin{equation}
  \frac{d N}{d \cos \Theta} = N_0 ( 1 + \alpha \, \cos \Theta),
\end{equation}
where $\Theta$ is the direction of the daughter baryon with respect to the polarisation
of the mother in the mother rest frame.
For e.g.\ $\xilamgam$, the decay asymmetry can then be measured by looking at the 
angle between the incoming $\Xi^0$ and the outgoing proton from the subsequent $\Lambda \to p \pi^-$ decay
in the $\Lambda$ rest frame (see Fig.~\ref{fig:Xi0Lamgamsketch})
Using this method, the measurement is independent of the unknown initial $\Xi^0$ polarisation.

\begin{figure}[htb]
  \begin{center}
    \epsfig{file=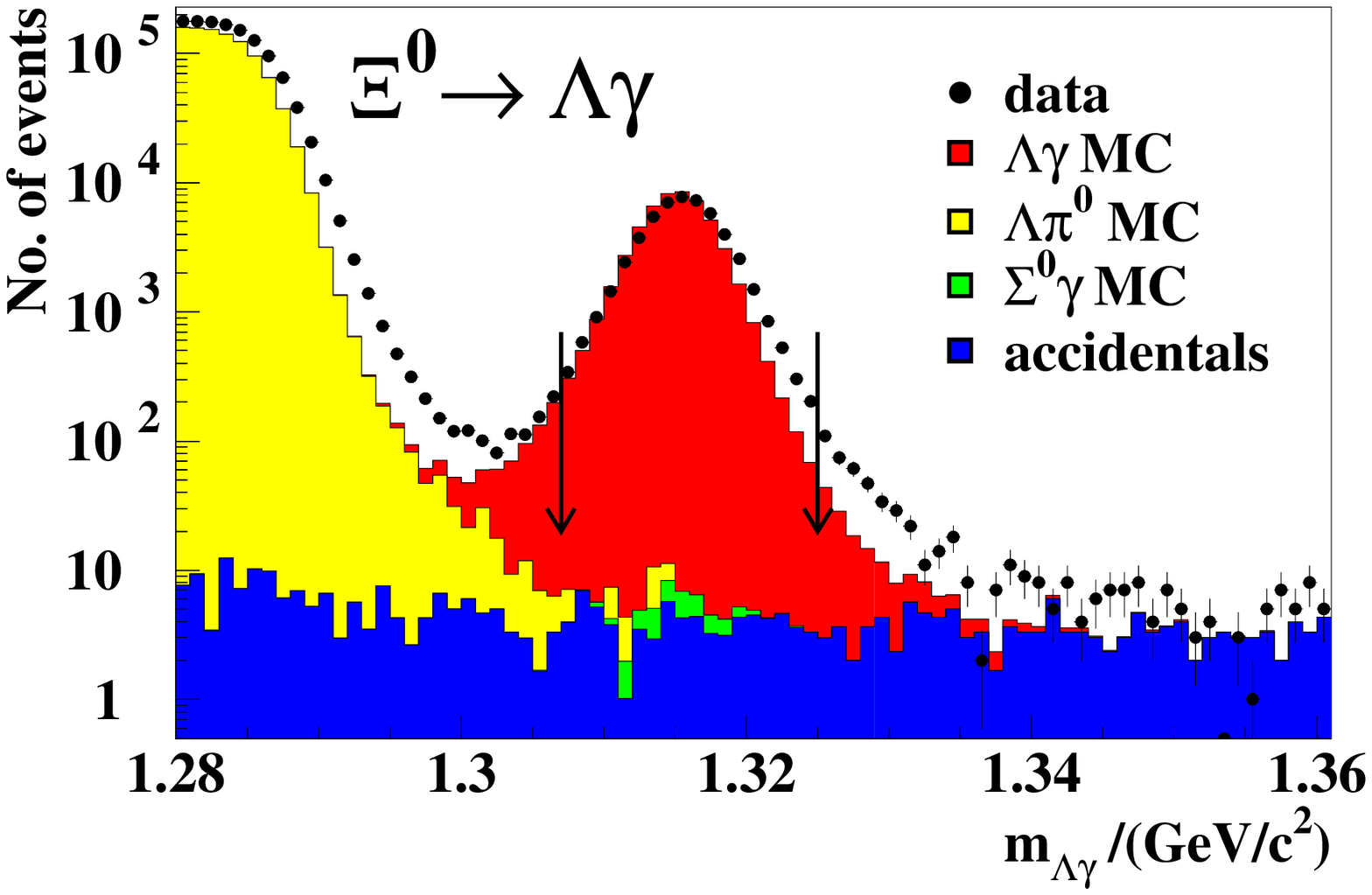,width=0.49\linewidth}
    \hspace*{\fill}
    \epsfig{file=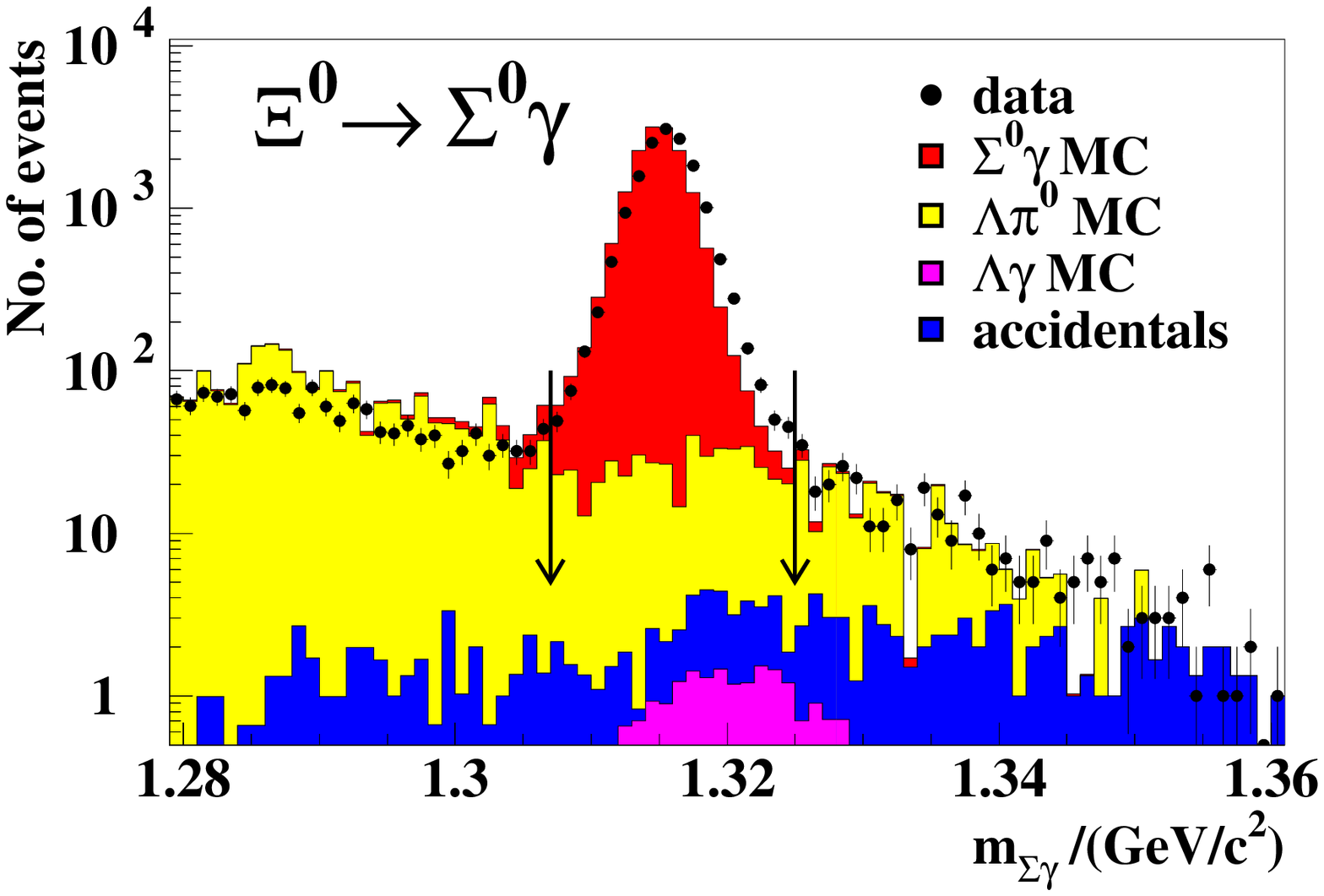,width=0.49\linewidth}
    \caption{$\xilamgam$ (left) and $\xisiggam$ (right) signal together with MC expectations for signal and backgrounds.}
    \label{fig:Xi0signals}
  \end{center}
\end{figure}

The NA48/1 experiment has selected 48314 $\xilamgam$ and 13068 $\xisiggam$ candidates (Fig.~\ref{fig:Xi0signals}).
The background contributions are $0.8\%$ for $\xilamgam$ and about $3\%$ for $\xisiggam$, respectively.

\begin{figure}[htb]
  \begin{center}
    \epsfig{file=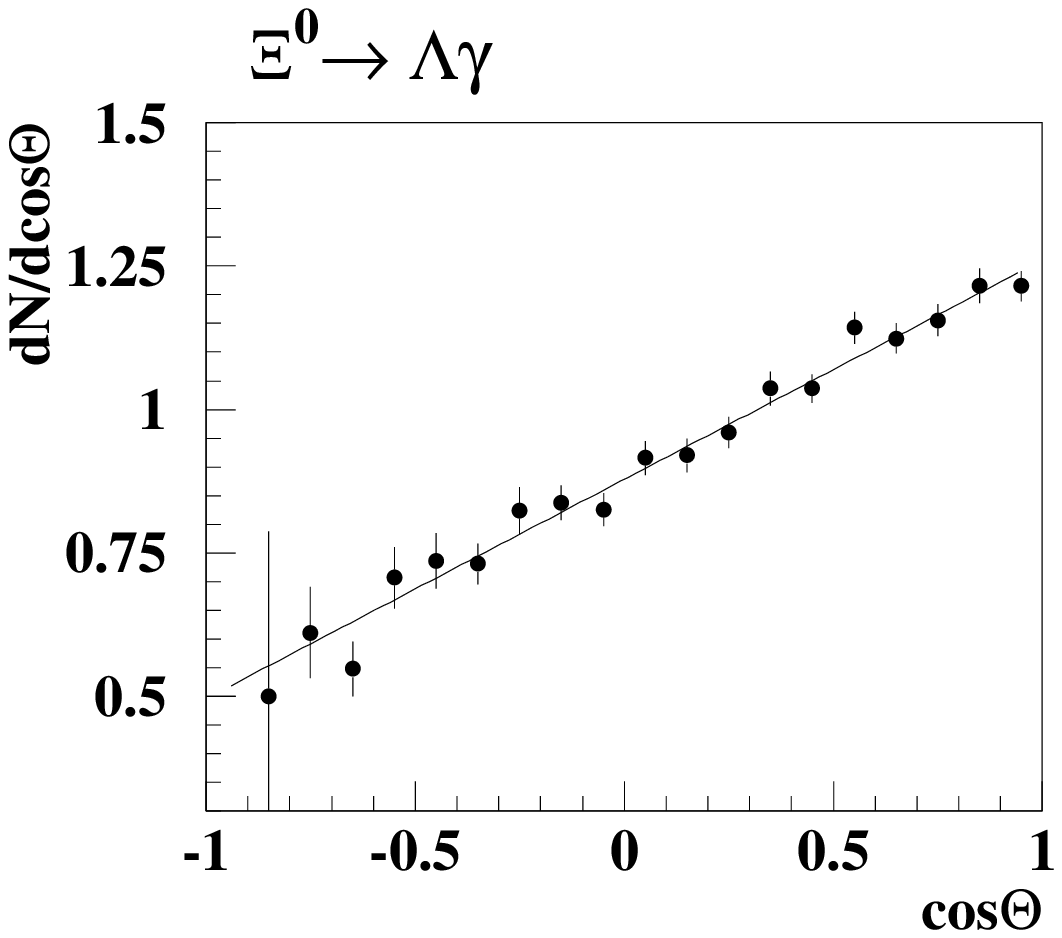,width=0.49\linewidth}
    \hspace*{\fill}
    \epsfig{file=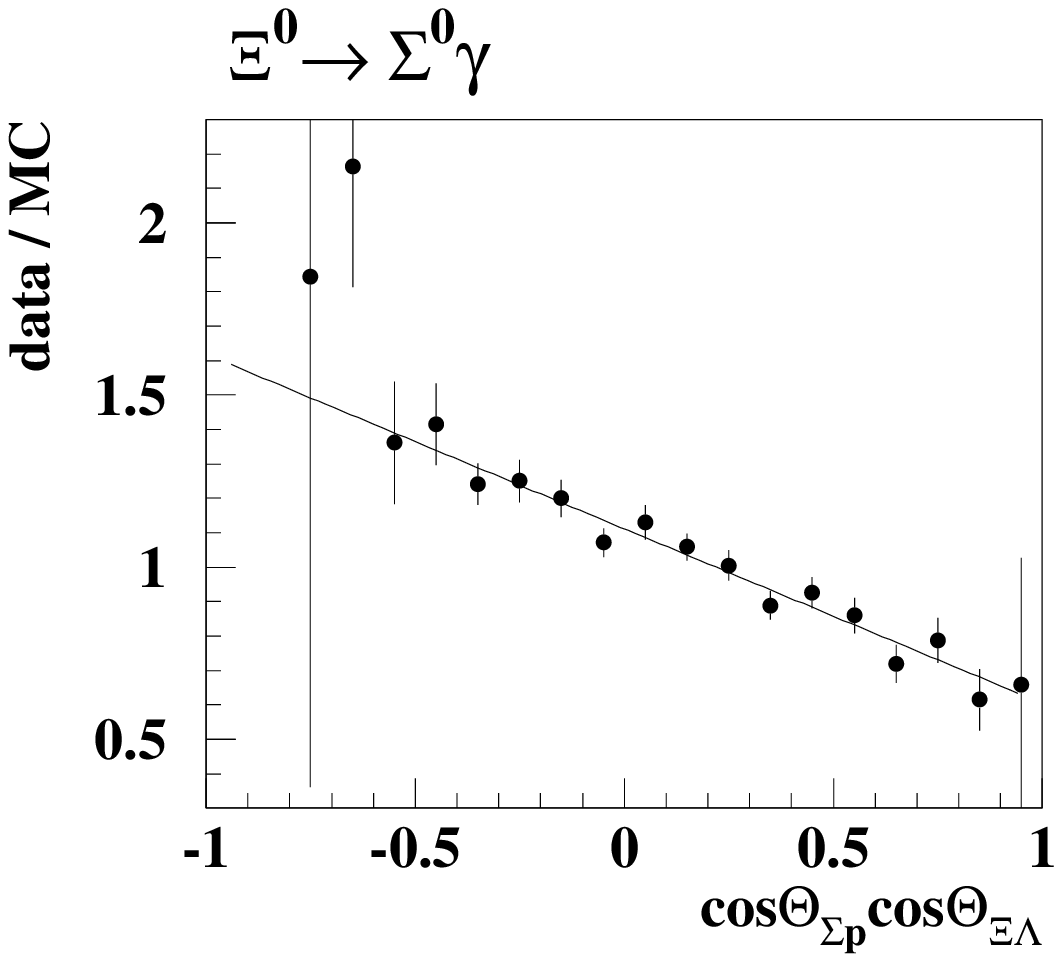,width=0.49\linewidth}
    \caption{Fits of the decay asymmetries in $\xilamgam$ (left) and $\xisiggam$ (right).}
    \label{fig:Xi0asymmetries}
  \end{center}
\end{figure}

Using these data, fits to the decay asymmetries have been performed. In case of $\xisiggam$, where
we have the subsequent decay $\Sigma^0 \to \Lambda \gamma$, the product 
$\cos \Theta_{\Xi \to \Sigma \gamma} \cdot \cos \Theta_{\Sigma \to \Lambda \gamma}$
has to be used for the fit.
Both fits show the expected linear behaviour on the angular parameters (Fig.~\ref {fig:Xi0asymmetries})

After correcting for the well-known asymmetry of $\Lambda \to p \pi^-$, values of
\begin{eqnarray}
  \alpha_{\Xi^0 \to \Lambda \gamma}  & = & -0.684 \pm 0.020 \pm 0.061  \quad \quad \text{and} \\
  \alpha_{\Xi^0 \to \Sigma^0 \gamma} & = & -0.682 \pm 0.031 \pm 0.065
\end{eqnarray}
are obtained,
where the first error is statistical and the second systematic.
These values agree with previous measurements by NA48 on $\xilamgam$~\cite{bib:Xi0Lamgam_NA48}
and KTeV on $\xisiggam$~\cite{bib:Xi0Siggam_KTeV}, but are much more precise.
In particular the result on $\xilamgam$ is of high theoretical interest, as it confirms the large
negative value of the decay asymmetry, which is difficult to accommodate for quark and vector meson dominance models.

\end{document}